\documentclass{egpubl}
\usepackage{egsgp18}

 \JournalSubmission    
 \electronicVersion 
 \usepackage{hyperref}

\ifpdf \usepackage[pdftex]{graphicx} \pdfcompresslevel=9
\else \usepackage[dvips]{graphicx} \fi

\PrintedOrElectronic

\usepackage{t1enc,dfadobe}

\usepackage{egweblnk}
\usepackage{cite}

\let\bf=\bfseries

\usepackage{microtype} 
\usepackage{breakcites}
\usepackage{ellipsis} 
\usepackage[mathletters]{ucs}
\usepackage[utf8x]{inputenc}
\usepackage{mathtools}
\usepackage{amsbsy}
\usepackage{upgreek}
\usepackage{dblfloatfix}

\usepackage[table]{xcolor}
\newcommand{\ra}[1]{\renewcommand{\arraystretch}{#1}}
\usepackage{tabularx}
\usepackage{booktabs}

\usepackage{amsmath}
\usepackage{amsfonts}
\usepackage{amssymb}

\definecolor{lightbluishgrey}{rgb}{0.78,0.86,0.93}

\newcommand{\newhl}[1]{#1}

\usepackage{layouts}

\usepackage{wrapfig}

\newcommand{\refequ}[1] {Equation~(\ref{equ:#1})}

\newcommand{\reffig}[1] {Figure~\ref{fig:#1}}

\makeatletter
\def\reffig{\@ifnextchar[{\@myreffigloc}{\@myreffignoloc}}
\def\@myreffigloc[#1]#2{Figure~\ref{fig:#2}, \emph{#1}}
\def\@myreffignoloc#1{Figure~\ref{fig:#1}}
\makeatother

\newcommand{\reftab}[1] {Table~\ref{tab:#1}}

\newcommand{\refsec}[1] {Section~\ref{sec:#1}}
\newcommand{\refapp}[1] {Appendix~\ref{app:#1}}

\let\mat = \mathbf
\usepackage{xfrac}

\usepackage{amsmath}
\usepackage{amssymb}    
\usepackage{cancel}
\usepackage[T1]{fontenc}

\newcommand{\R}{\mathbb{R}}

\newcommand{\vc}[1]{\mathbf{#1}}
\newcommand{\C}{\mat{C}}

\newcommand{\transpose}{{\mathsf T}}

\renewcommand{\a}{\vc{a}}

\newcommand{\n}{\vc{n}}

\renewcommand{\u}{\vc{u}}
\newcommand{\vv}{\vc{v}}

\newcommand{\x}{\vc{x}}
\newcommand{\y}{\vc{y}}
\newcommand{\z}{\vc{z}}
\newcommand{\A}{\mat{A}}
\newcommand{\B}{\mat{B}}

\newcommand{\G}{\mat{G}}

\renewcommand{\L}{\mat{L}}
\newcommand{\M}{\mat{M}}

\newcommand{\T}{\mathcal{T}}

\newcommand{\V}{\mat{V}}

\usepackage{siunitx}
\usepackage{graphicx}

\usepackage{wrapfig}
\newcommand{\wf}[6]{%
  \begin{wrapfigure}[#1]{#2}{#3}%
  \centering%
  \includegraphics[#4]{#5}
  #6%
\end{wrapfigure}}

\makeatletter
\let\save@mathaccent\mathaccent
\newcommand*\if@single[3]{%
  \setbox0\hbox{${\mathaccent"0362{#1}}^H$}%
  \setbox2\hbox{${\mathaccent"0362{\kern0pt#1}}^H$}%
  \ifdim\ht0=\ht2 #3\else #2\fi
  }
\newcommand*\rel@kern[1]{\kern#1\dimexpr\macc@kerna}
\newcommand*\widebar[1]{\@ifnextchar^{{\wide@bar{#1}{0}}}{\wide@bar{#1}{1}}}
\newcommand*\wide@bar[2]{\if@single{#1}{\wide@bar@{#1}{#2}{1}}{\wide@bar@{#1}{#2}{2}}}
\newcommand*\wide@bar@[3]{%
  \begingroup
  \def\mathaccent##1##2{%
    \let\mathaccent\save@mathaccent
    \if#32 \let\macc@nucleus\first@char \fi
    \setbox\z@\hbox{$\macc@style{\macc@nucleus}_{}$}%
    \setbox\tw@\hbox{$\macc@style{\macc@nucleus}{}_{}$}%
    \dimen@\wd\tw@
    \advance\dimen@-\wd\z@
    \divide\dimen@ 3
    \@tempdima\wd\tw@
    \advance\@tempdima-\scriptspace
    \divide\@tempdima 10
    \advance\dimen@-\@tempdima
    \ifdim\dimen@>\z@ \dimen@0pt\fi
    \rel@kern{0.6}\kern-\dimen@
    \if#31
      \overline{\rel@kern{-0.6}\kern\dimen@\macc@nucleus\rel@kern{0.4}\kern\dimen@}%
      \advance\dimen@0.4\dimexpr\macc@kerna
      \let\final@kern#2%
      \ifdim\dimen@<\z@ \let\final@kern1\fi
      \if\final@kern1 \kern-\dimen@\fi
    \else
      \overline{\rel@kern{-0.6}\kern\dimen@#1}%
    \fi
  }%
  \macc@depth\@ne
  \let\math@bgroup\@empty \let\math@egroup\macc@set@skewchar
  \mathsurround\z@ \frozen@everymath{\mathgroup\macc@group\relax}%
  \macc@set@skewchar\relax
  \let\mathaccentV\macc@nested@a
  \if#31
    \macc@nested@a\relax111{#1}%
  \else
    \def\gobble@till@marker##1\endmarker{}%
    \futurelet\first@char\gobble@till@marker#1\endmarker
    \ifcat\noexpand\first@char A\else
      \def\first@char{}%
    \fi
    \macc@nested@a\relax111{\first@char}%
  \fi
  \endgroup
}
\makeatother

\usepackage[ruled,vlined]{algorithm2e}
\usepackage{etoolbox}
\usepackage{array}

\newcommand{\figs}{}
\def\figs/{figs/}

\newcommand{\dA}{\;dA}

\let\min\relax
\DeclareMathOperator*{\min}{\text{min }}

\usepackage[verbose]{newunicodechar}

\newcommand{\diag}[1]{\text{diag}\left(#1\right)}
\newcommand{\blkdiag}[1]{\text{blkdiag}\left(#1\right)}

\begin{document}

\title{\noindent
  \vspace*{-0.1cm}
\newhl{Solid~Geometry~Processing~on~Deconstructed~Domains}
  \vspace*{-0.2cm}
}
\author[Sellán et al.]{
  Silvia Sellán¹ $\quad$
  Herng Yi Cheng² $\quad$
  Yuming Ma³ $\quad$
  Mitchell Dembowski⁴ $\quad$
  Alec Jacobson³\\
  ¹University of Oviedo $\quad$
  ²Massachusetts Institute of Technology $\quad$
  ³University of Toronto $\quad$
  ⁴Ryerson University
  \vspace*{-0.1cm}
}

\maketitle
\begin{abstract}
Many tasks in geometry processing are modeled as \newhl{variational problems}
solved numerically using the finite element method.
For solid shapes, this requires a volumetric discretization, such as a boundary
conforming tetrahedral mesh.
Unfortunately, tetrahedral meshing remains an open challenge and existing
methods either struggle to conform to complex boundary surfaces or require
manual intervention to prevent failure.
Rather than create a single volumetric mesh for the entire shape, we advocate
for solid geometry processing on \emph{deconstructed domains}, where a large
and complex shape is composed of overlapping solid subdomains.
As each smaller and simpler part is now easier to tetrahedralize, the question
becomes how to account for overlaps during problem modeling and how to couple
solutions on each subdomain together \emph{algebraically}.
We explore how and why previous coupling methods fail, and propose a method that
couples solid domains only along their boundary surfaces. 
We demonstrate the superiority of this method through empirical convergence
tests and qualitative applications to solid geometry processing on a variety
of popular second-order and fourth-order partial differential equations.
\end{abstract}

\begin{figure}
  \includegraphics[width=\linewidth]{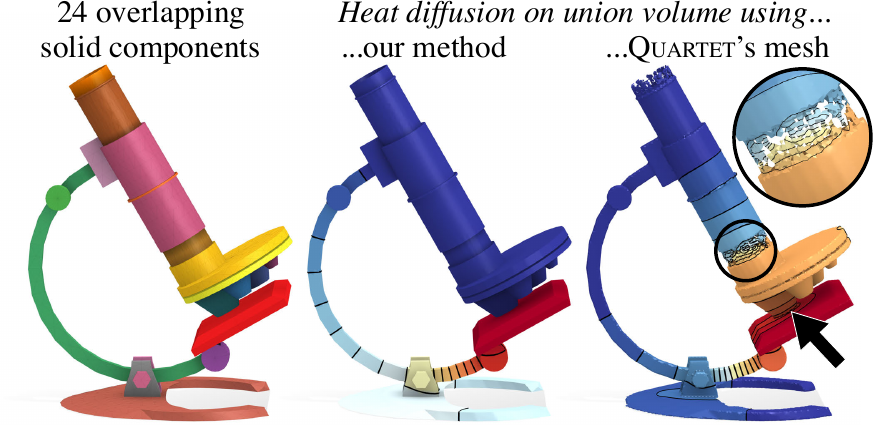}
\caption{
\newhl{
  \textsc{Tetgen} fails to tetrahedralize this Microscope, even after
  preprocessing the input using \cite{Zhou:2016}. \textsc{Quartet} successfully
  outputs a tet mesh, but thin parts are poorly approximated (inset) and close
  features are merged (arrow).}
}
\label{fig:microscope}
\end{figure}

\section{Introduction}
\label{sec:intro}
%
Many tasks in computer graphics and geometry processing can be modeled
mathematically as solutions to partial differential equations (PDEs) over a
compact spatial domain.
For example, shape-aware scattered data interpolation can be modeled as a
solution to the Laplace equation ($∆u = 0$).
Smooth detail-preserving shape deformations can be efficiently parametrized
using solutions to a bi-Laplace equation ($∆²u = 0$).
Even computation of geodesic distances can be captured via iterative solutions
to a Poisson equation ($∆u = f$).
These applications --- and many others --- rely on \emph{discretization} to
realize their solutions on the complex shapes found throughout computer
graphics.
The most common discretization is via the finite-element method (FEM) using
piecewise-linear functions defined over a simplicial mesh.
For problems over solid regions in $\R^3$, this typically requires constructing
a tetrahedral mesh that fills the volume bounded by a given surface.
Compared to regular grids, unstructured tetrahedral meshes afford spatially
varying resolution and complex boundary surfaces --- \emph{in theory,
at least}.

In practice, constructing tetrahedral meshes is a fragile process.
While the application of linear FEM is often straightforward after posing a
problem in the smooth setting, the actual creation of a valid tetrahedral mesh
inside a triangle mesh is often left to an \emph{ad hoc} patchwork of heuristics
including manual intervention and mesh repair.
Existing automatic meshing methods fall short.  They either fail too often,
create too poor quality elements, or approximate too loosely the input domain
boundary\newhl{, as shown in \reffig{microscope}}. 

\begin{figure}
  \includegraphics[width=\linewidth]{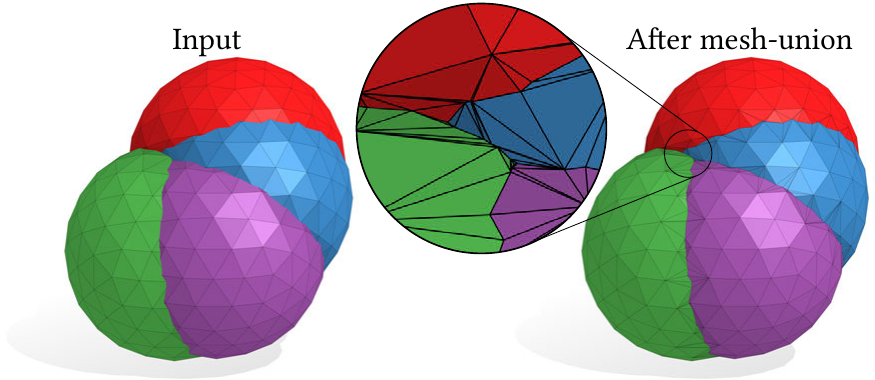}
  \caption{\label{fig:bad-spheres} Even if all input meshes are ``clean'' (e.g.,
  four overlapping geodesic spheres on left), their exact mesh-union may have
  arbitrarily poor quality triangles.  These in turn trip up conforming Delaunay
  tet-meshers: e.g.,\ TetGen \protect\cite{tetgen} fails on this example.}
\end{figure}

\wf{}{r}{1.5in}{trim={50mm 0mm 0mm 30mm},width=\linewidth}{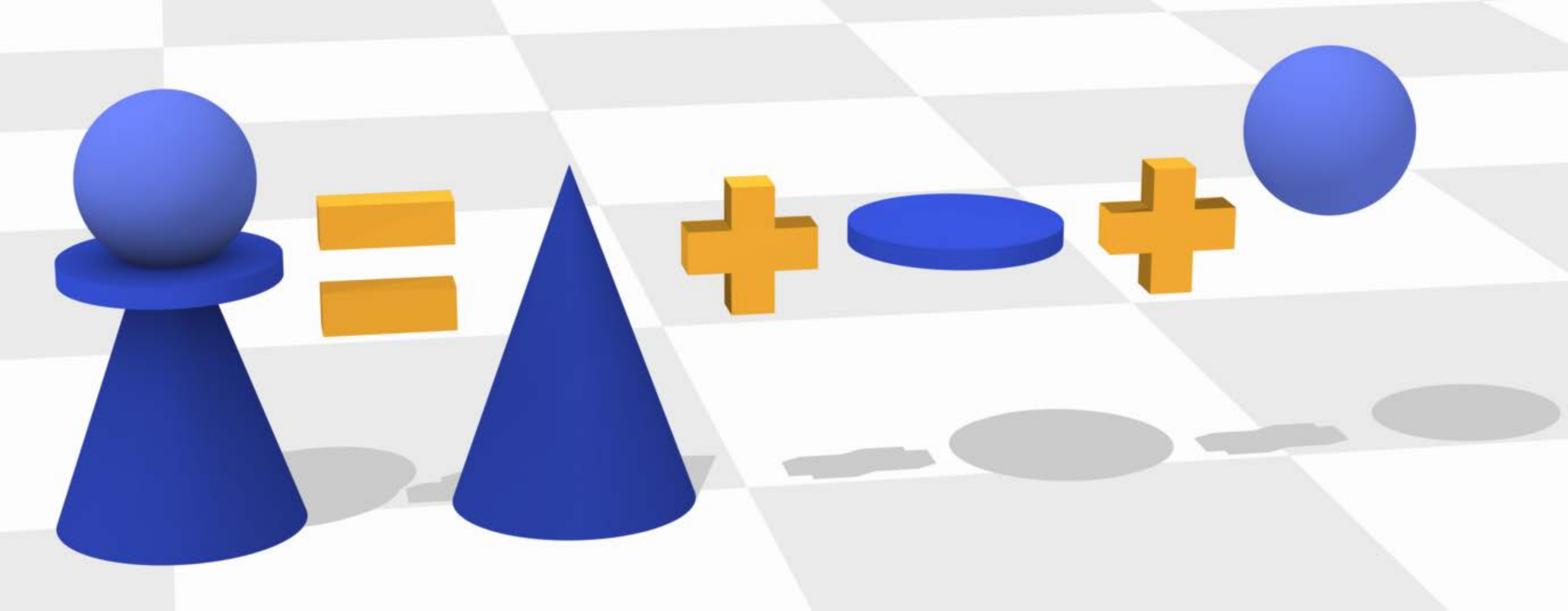}{}
We consider an interesting class of shapes that are --- at least
conceptually if not literally --- described as the union of simpler domains (see
inset). 
The traditional conforming tetrahedralization pipeline would proceed by first
computing the result of a surface mesh union operation and then attempt to mesh
the interior.
However, even if the input triangle meshes are ``clean'', the exact mesh boolean
result may be host to a number of issues that trip up available
tetrahedralization heuristics: fine features, small voids, and poorly shaped
elements (see \reffig{bad-spheres}).
In general, the \emph{exact} result introduces many new vertices \newhl{whose
coordinates are rational numbers.
 Naively rounding such vertices to
floating-point coordinates may introduce self-intersections, and efficient
rounding while preventing intersections in 3D is still an open problem
\cite{Fortune97}.}

In this paper, we propose an alternative to this error-prone pipeline.
We consider inputs as \emph{deconstructed domains}, composed as the union
of any number of simpler shapes.
We tetrahedralize subdomains independently with no requirement to share vertex
positions or combinatorics.
Subdomains are treated democratically, without a priority ordering or hierarchy.
We then couple discrete PDEs or \newhl{variational} problems defined on each
domain \emph{algebraically}.

This coupling requires care. 
The given PDE or \newhl{variational} problem must be adapted for overlapping
domains to avoid bias or double-counting in twice-covered regions.
Via null-space analysis, we show that naive equality constraints leads to
\emph{locking}, artificial error that does not vanish under resolution
refinement.

\newhl{We borrow ideas from domain decomposition and immersed boundary methods
to}
derive a general-purpose boundary-only coupling in the smooth setting and
then demonstrate its effectiveness for linear FEM discretizations of common
problems in 1D, 2D, and 3D (e.g., Poisson equations).
\newhl{Domains are coupled with hard constraints resulting in a parameterless
method.}
Further, we extend our results to boundary-only higher-order coupling for
applications of mixed FEM for fourth-order problems (e.g., bi-Laplace
equation).

\section{Related Work}
\label{sec:related}
Our goal is to improve the robustness of geometry processing that requires
solving partial differential equations (PDEs) on solid shapes.

\paragraph*{PDE Solvers in Geometry Processing}
Improving the accuracy, robustness and performance of solvers for geometry
problems is a core area of interest
\cite{BotschBK05,Krishnan:2013,Kazhdan:2015:FEP,Devito:2017,Herholz:2017:LSS,Shtengel:2017}.
While we focus on robustness with respect to the input representation, our work
relates to these as an \emph{algebraic} preprocess or filter on the eventual
linear system or optimization problem.
This is in contrast to geometric approaches to robustness such as remeshing
\cite{BotschKobbelt04,Sokolov:2016:HM,Gao:2017:RHM} or enclosing a shape in a
cage \cite{HarmonicCoordinates07,Sacht:2015:NC}.
\newhl{For example, using the boundary element method avoids volumetric meshing
altogether (e.g., \cite{James:1999:AAR,dhbwg16,solomon2017boundary}), however,
this also limits the class of problems that can be solved.}
In contrast to methods specific to one application (e.g., character
skinning \cite{Bharaj:2012:AutoRig}), our method applies to a general class of
PDEs.

\paragraph*{Constructive Solid Geometry}
Emerging technologies such as 3D printing and virtual reality have ignited
broader interest in geometric modeling.
The result is that we have a huge amount of geometric data, but that data is
rarely composed of a single, watertight, non-self-intersecting, oriented
manifold surface \cite{thingi10k}.
Instead, people create using constructive solid geometry tools 
like \textsc{OpenSCAD} and \textsc{Tinkercad} that happily allow overlapping
simpler models to create a larger, more complex shape.
%
Early digital constructive solid geometry complements this modeling paradigm
with fast evaluation using implicit functions (e.g., \cite{Wyvill:1986}) and 
GPU-friendly rendering \cite{Goldfeather:1986:FCG}.
While the complexity of available meshed surface geometry grows, researchers
have devised interesting and interactive ways to create complex models using
preexisting detailed parts \cite{Gal2006,Chaudhuri:2010:DSC,Chaudhuri2011}.

Most volumetric solvers do not consider the upstream modeling process and
instead require a single volumetric mesh as input. This puts a heavy burden on
modeling tools to maintain a clean surface geometry via mesh ``surgery''
operations \cite{Sharf:2006:SnapPaste,schmidt2010drag,schmidt2010meshmixer}.
Despite recent progress on robust boolean operations for triangle meshes
\cite{Bernstein:2009:FEL,Barki2015,Zhou:2016}, the resulting meshes may have
arbitrarily poor aspect ratio (see \reffig{bad-spheres}) preventing or damaging
tetrahedralization.
In contrast, we operate directly on the overlapping subdomain representation
common to solid modeling.

\paragraph*{Tetrahedral Meshing}
Although our work is an effort to subvert tetrahedral meshing and its
issues, we still rely heavily on the progress and open source software from
this literature.
The fundamental challenge at the core of tetrahedral meshing is the balance
between ensuring high-quality elements (see, e.g., \cite{Shewchuk02}) and
conforming to a given input surface.
While a complete meshing survey is outside the scope of this paper, we identify
issues and previous works as they relate to our setting.

\begin{figure}[b!]
\includegraphics[width=\linewidth]{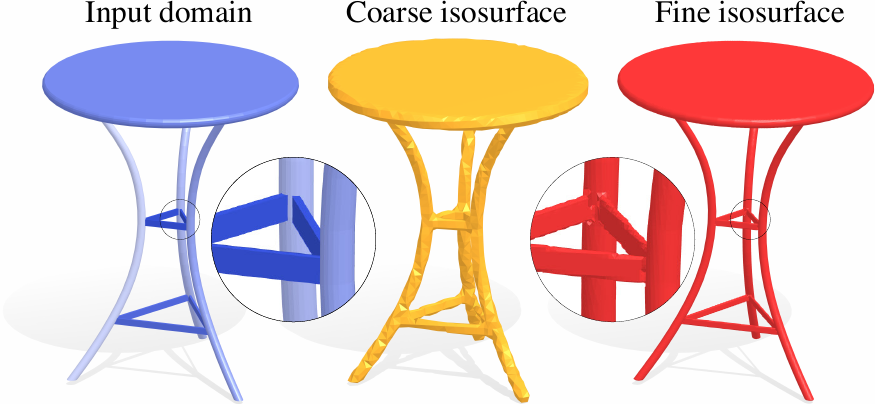}
\caption{
\label{fig:table}
This seemingly innocuous union (left) causes the conforming Delaunay method
\textsc{TetGen} \protect\cite{tetgen} to fail.
Implicit or query-based methods have trouble capturing fine structures (middle)
without resorting to high resolutions (right).}
\end{figure}

Conforming Delaunay Tetrahedralization methods maintain the input geometry (and
combinatorics) exactly by inserting input vertices and faces into a Delaunay
triangulation and then improving element quality via local operations and
additional Steiner vertices \cite{cheng2012delaunay}.
In practice, the software \textsc{TetGen} \cite{tetgen} implements many
state-of-the-art algorithms and heuristics. 
The success rate of \textsc{TetGen} is not 100\% (see \reffig{table}), but it
succeeds far more often when inputs are smaller and simpler, without spatially
close parts or small triangles.
We use \textsc{TetGen} in most of our examples, but run it on each subdomain,
rather than the complex, complete shape.

Alternatively, other meshing methods work by employing a background grid
\cite{labelle2007isosurface} or implicit representation of the input shape
\cite{alliez2005variational}.
These methods ensure good quality elements by construction, but struggle to
closely approximate the input shape geometry --- especially in the presence of
sharp features.
Doran et al.\ \shortcite{Doran:2013} provide an open source implementation,
\textsc{Quartet} and while robust in the sense of successfully outputting a
mesh, this method will join together close features and fail to resolve thin
parts (see \reffig{microscope}).
In contrast to Cuilliere et al.\ \cite{CuilliereFD12}, we avoid computing a
unified mesh and do not require matching or correspondence between vertices or
combinatorics of overlapping meshes.

\paragraph*{Domain Decomposition}
The idea of coupling solutions to partial differential equations across
overlapping domains is quite old \cite{Schwarz1870} and well studied.
The majority of previous methods for overset and non-matching grids focus on
domain decomposition for parallel, offline computation using iterative solvers
\cite{Smith2004domain}.
\newhl{Alternatively, \emph{immersed boundary methods} \cite{immersedPeskin} use
similar constraints to couple the simulation of one or many objects embedded in
a background simulation. For example, coupling a floating elastic body to a
fluid simulation (e.g., \cite{Guendelman:2005}).}

In contrast, our interest is in reducing the burden of tetrahedralization while
maintaining the complexity of shapes found in graphics and geometry
processing.
We treat coupling as a hard constraint to single linear system solved using
modern, large sparse linear solvers (e.g., \cite{cholmod3,andersen17}).
\newhl{No sub-domain has preference over another.}


\newhl{English et al.}\ \shortcite{English:2013:CGW} simulate
water at varying resolutions by allowing regular finite-difference grids to
rigidly overlap. Their method assigns priorities to grids and stitches
higher-priority grids along their boundaries into lower-priority grids to solve
a Poisson equation. Similar so-called \emph{Chimera grids} \cite{Benek1985} are
found in early fluid simulations on comparatively simple domains
\cite{Benek1983,henshaw1994fourth,Kiris1997,dobashi2008fast}.
Henshaw describes how boundary values of one grid are interpolated using ghost
points. This method is applied to overlapping regular Cartesian or polar grids.
Malgat et al.\ \shortcite{Malgat:2015:MHM} couple overlapping discretizations
for elasticity simulation via energy minimization. Their method requires a
hierarchical ordering.

Overset grid methods (e.g.,
\cite{Nakahashi1999,Loehner,bercovier2015overlapping}) often assume that the
domain has been designed with an overlapping solver in mind.
High resolution grids near important areas naturally have well defined and known
priority over coarse background grids. Instead, we consider the case where
subdomain priorities are not known and domains merely serve as an overlapping
subdivision. The resolution of a single grid may itself be adaptive.

Schwarz domain decomposition can be interpreted in the context of discontinuous
Galerkin finite element method (DGFEM) \newhl{or extended FEM (XFEM)
\cite{KaufmannThesis}}, where subdomains are interpreted as large, high-degree
elements and coupling is analogous to interface conditions.
Edwards \& Bridson \shortcite{EdwardsB15} propose such a solver for Poisson,
elasticity and bi-Laplace problems. Their overlapping subdomains are extracted
from a unified grid of the entire domain.


\section{Smooth Foundations}
\label{sec:smooth-beginnings}
We first consider a partial differential equation (PDE) involving a smooth
function $u$ defined over a volumetric (i.e., co-dimension zero) domain $Ω ⊂ \R^d$
with appropriate boundary conditions applied to $u$ on the boundary of the
domain $∂Ω$.
We focus specifically on elliptic PDEs resulting from energy minimizations
common in geometry processing. For example, minimizing the squared gradient
(i.e., Dirichlet energy) minus a unit potential, subject to fixing the value of
$u$ to a known function $g$ on the boundary of the domain,
\begin{align}
  \label{equ:smooth-dirichlet-energy}
  \min_u                     & ∫_Ω \left( \tfrac{1}{2}\|∇u\|² - u \right)\dA  \\ 
         \text{subject to } & u(\x) = g(\x) & ∀ \x ∈ ∂Ω,
\end{align}
results in the second-order Poisson equation on the interior,
\begin{align}
  \label{equ:smooth-poisson-equation}
  ∆u &= 1 & ∀ \x ∈ Ω.
\end{align}

\begin{figure}[b!]
  \includegraphics[width=\linewidth]{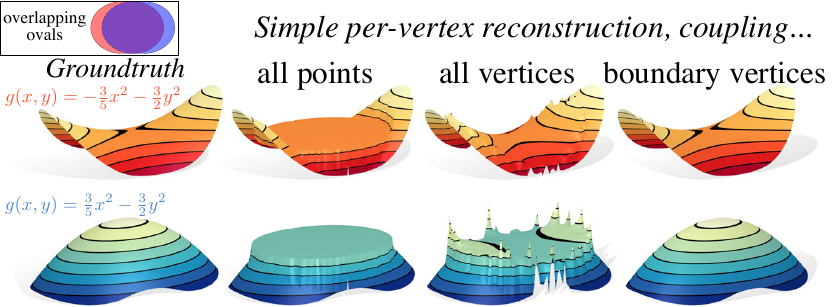}
  \caption{\label{fig:saddle-vs-parabola}
  The constraint space is unrelated to the energy.
  Consider a simple reconstruction energy $\min_u ∫_{Ω₁ ∪ Ω₂} (u-g)² \dA$.
  Constraining \emph{all points} in the overlap locks to a
  linear function; \emph{all vertices} looks promising for saddle-shaped
  functions, but cannot reproduce positive curvature functions;
  \emph{boundary vertices} avoids locking.}
\end{figure}

\begin{figure*}
\includegraphics[width=\linewidth]{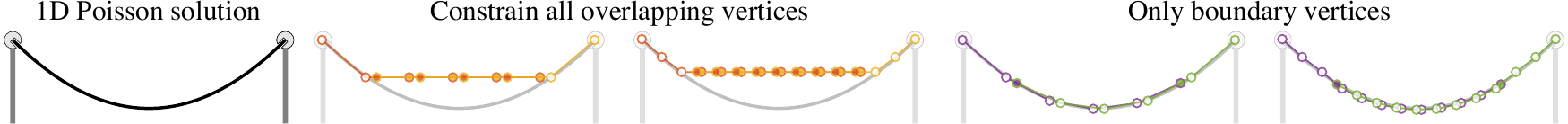}
\caption{\label{fig:1d-poisson} 
  Left-to-right: the Poisson solution in 1D is a simple parabola, easily
  approximated using linear finite element method.
  If the discrete domain is given as two overlapping meshes (red \& yellow), 
  then enforcing equality at \emph{all vertices} in the overlap results in the
  solution \emph{locking} to a linear function. This problem does not go away
  with mesh refinement. Coupling the meshes only at the sub-domain boundary
  (purple \& green)
  alleviates locking, and converges with refinement.}
  \includegraphics[width=\linewidth]{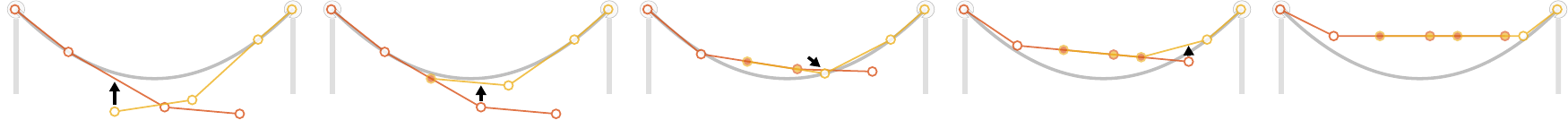}
  \caption{\label{fig:1d-domino-effect} Locking occurs at any scale and can be
  understand as a ``Domino effect'' when enforcing constraints one-by-one. In
  1D, consecutive constraints force the solution to a line: eventually the
  entire overlapping region must be a single line.}
  \vspace*{-0.4cm}
\end{figure*}

\wf{8}{r}{1.5in}{trim={5mm 0mm 0mm 3mm},width=\linewidth}{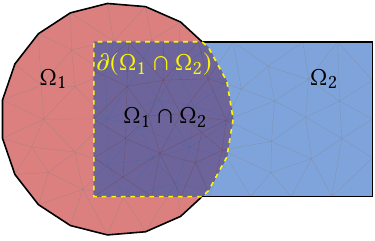}{}
Suppose we are incapable of measuring an energy directly over all of the domain
$Ω$, but instead are only able to measure energies over two \emph{overlapping}
subdomains $Ω₁,Ω₂⊂\R^d$ whose union composes the original domain $Ω₁ ∪ Ω₂ = Ω$.
By replacing $u$ with 
new $u₁$ and $u₂$ over each respective subdomain, we can write 
the original minimization problem in
\refequ{smooth-dirichlet-energy}, breaking the integral into the non-overlapping
parts in each subdomain ($Ω₁ \setminus Ω₂$ and $Ω₂ \setminus Ω₁$) and their
intersection ($Ω₁∩Ω₂$)
and adding a \emph{pointwise} equality coupling constraint,
\begin{align}
  \label{equ:smooth-deconstructed-energy}
  \min_{u₁,u₂}               &\hphantom{\frac{1}{2}}∫_{Ω₁ \setminus Ω₂} \left( \tfrac{1}{2}\|∇u₁\|² - u₁ \right) \dA+\\
                           &+\hphantom{\frac{1}{2}}∫_{Ω₂ \setminus Ω₁} \left( \tfrac{1}{2}\|∇u₂\|² - u₂ \right) \dA+\\
                            \label{equ:one-half}
                            &+\frac{1}{2}∫_{Ω₁ ∩ Ω₂}         \left( \tfrac{1}{2}\|∇u₁\|² - u₁ +  \tfrac{1}{2}\|∇u₂\|² - u₂ \right) \dA
\end{align}
\begin{align}
  \label{equ:smooth-dirichlet-1}
         \text{subject to } & u₁(\x) = g(\x) & ∀ \x ∈ ∂Ω∩∂Ω₁, \\
  \label{equ:smooth-dirichlet-2}
         \text{and }        & u₂(\x) = g(\x) & ∀ \x ∈ ∂Ω∩∂Ω₂, \\
         \label{equ:smooth-pointwise-constraint}
         \text{and }        & u₁(\x) = u₂(\x) &∀ \x ∈ Ω₁∩Ω₂.
\end{align}

\paragraph*{Advantage of working with energies}
The appearance of the $½$ factor before the integrated energy in the
intersection $Ω₁ ∩ Ω₂$ region (see \refequ{one-half}) would not be so obvious if
we had worked with the Poisson problem directly as a PDE (see
\refequ{smooth-poisson-equation}).
However, viewed as \newhl{variational problem}, the necessity of the $\sfrac{1}{2}$ is
clear: we should not double count the energy contributed in this region.
%

\section{Discrete Locking}
The deconstructed energy optimization problem in
\refequ{smooth-deconstructed-energy} only involves first derivatives and linear
equality constraints. It is tempting to jump to a finite element method (FEM) discretization
using piecewise-linear elements for each subdomain $Ω₁$ and $Ω₂$, e.g., hat
functions (over polylines for $d=1$, triangle meshes for $d=2$ and tetrahedral
meshes in $d=3$) 
\begin{equation}
  u_i(\x) = ∑_{j=1}^n  u_{ij} φ_{ij}(\x),
\end{equation}
with interpolated values at the $n$ vertices given as a vector
$\u_i ∈ \R^n$.

If the meshes over $Ω₁$ and $Ω₂$ have \emph{only and exactly} \newhl{coincident} 
vertices \emph{and} compatible combinatorics in the intersection region $Ω₁∩Ω₂$, then
we call them \emph{matching}. In this special case, enforcing the point-wise equality
constraint in \refequ{smooth-pointwise-constraint} is equivalent to merging
the meshes. The solution search space is exactly as rich as linear FEM over the
merged mesh.

For meshes with vertices in general position, perfect coincidence never happens.
Pointwise equality immediately reduces the search space to piecewise linear
functions that exist mutually in both linear FEM function spaces over the
intersection $Ω₁∩Ω₂$.
In the general \emph{non-matching} case, the constraint reduces the search space
dramatically: \emph{only} functions that take on a linear function over $Ω₁∩ Ω₂$
remain (see \reffig{saddle-vs-parabola}, left).

This is an extreme case of what is known as \emph{locking} in the FEM
literature \cite{zienkiewicz2000fem}.
Locking is an artificial stiffening of system
during discretization.
In our case, the constraints are so strict that
only rather boring functions remain.
These functions can be arbitrarily far from the desired solution and
discretization refinement by adding more (general position) vertices will not
help (see \reffig{1d-poisson}).

\subsection{\newhl{Constraints at All Vertices Causes Locking}}
One immediate strategy 
is to require equality only at mesh vertices.
This ties the values at one mesh's vertices to the piecewise linearly
interpolated value on the other mesh via a linear equality constraint and
\emph{vice-versa}, e.g.,:
\begin{align}
  \label{equ:naive-constraints}
  u_{1i} &= \sum\limits_{j=1}^{n₂} u_{2j} φ_{2j}(\vv_{1i}) & ∀ i \text{ such that } \vv_{1i} ∈ Ω₁∩Ω₂, \\
  u_{2j} &= \sum\limits_{i=1}^{n₁} u_{1i} φ_{1i}(\vv_{2j}) & ∀ j \text{ such that } \vv_{2j} ∈ Ω₂∩Ω₁,
\end{align}
where $\vv_{ki} ∈ \R^d$ is the position of the $i$th vertex in the mesh over
subdomain $Ω_k$.
The coefficients obtained by evaluating the hat functions $φ_{\ell j}(\vv_{ki})$
of the \emph{other} mesh over subdomain $Ω_\ell$ are simply the barycentric
coordinates of $\vv_{ki}$ in the containing simplex (e.g., tetrahedron for
$d=3$).

We can collect these constraints in matrix form
\begin{equation}
  \label{equ:constraint-matrix}
  \C \left(\begin{array}{c} \u₁ \\ \u₂ \end{array}\right) = \mathbf{0},
\end{equation}
where $\C ∈ \R^{(m₁ + m₂) × (n₁ + n₂)}$ is a sparse rectangular matrix, where
each row corresponds to one of the $m₁$ vertices of the mesh over $Ω₁$ lying in
$Ω₂$ or \emph{vice-versa}.

These linear equality constraints are easy to implement in
practice (e.g., via the null space or Lagrange multiplier method).
Unfortunately, these constraints do not alleviate locking.

\begin{figure}
  \includegraphics[width=\linewidth]{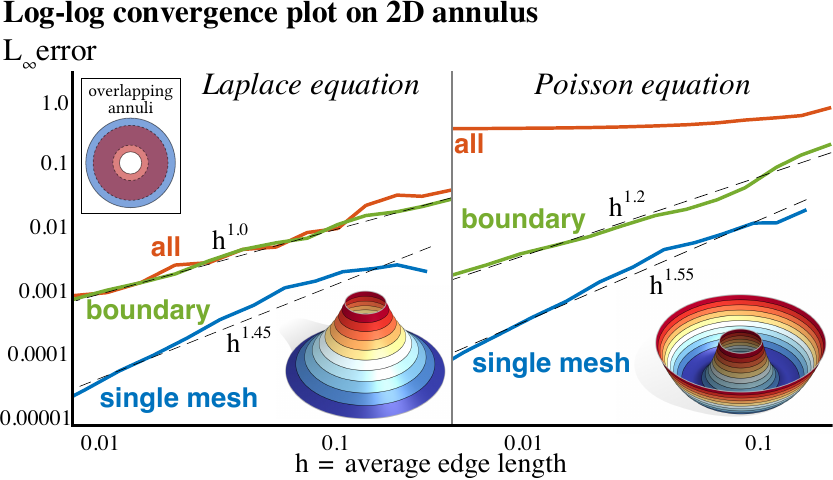}
  \caption{\label{fig:2d-annulus-convergence}
    Constraining \emph{all} vertices in the overlap between two concentric
    annuli matches the convergence rate of our \emph{boundary only}
    constraints for $∆z = 0$ (left). However, constraining
    \emph{all} does not converge for 
    $∆z = 1$ (right), while \emph{boundary only} maintains
    convergence. For reference: a non-overlapping, \emph{single mesh} converges.
  }
  \vspace*{-0.2cm}
\end{figure}

In $\R¹$, constraining all vertices in the overlapping region $Ω₁∩Ω₂$ is
catastrophic.
Intuitively, if a segment of the mesh over $Ω₁$ overlaps with a segment of the
mesh over $Ω₂$ then both pairs of vertices will have to lie on the same line. In
the worst case, an alternating order of vertices from $Ω₁$ and $Ω₂$ creates a
\emph{domino effect}, and the entire intersection region locks to the same
linear function (see \reffig{1d-domino-effect}).

It is tempting to extrapolate that these constraints will always result in
point-wise locking, but
in higher dimensions ($d>1$), locking from vertex constraints is
more nuanced.
We observe in \reffig{saddle-vs-parabola} that the constraint space
created by coupling all vertices struggles to reproduce a round parabolic
function and more easily reproduces a saddle-shaped hyperbolic function.
Indeed, imposing this constraint when solving a Laplace equation (saddle-shaped
solution) we see significantly better convergence with respect to mesh
resolution than when solving a parabolic Poisson equation (see
\reffig{2d-annulus-convergence}).

\begin{figure}
\includegraphics[width=\linewidth]{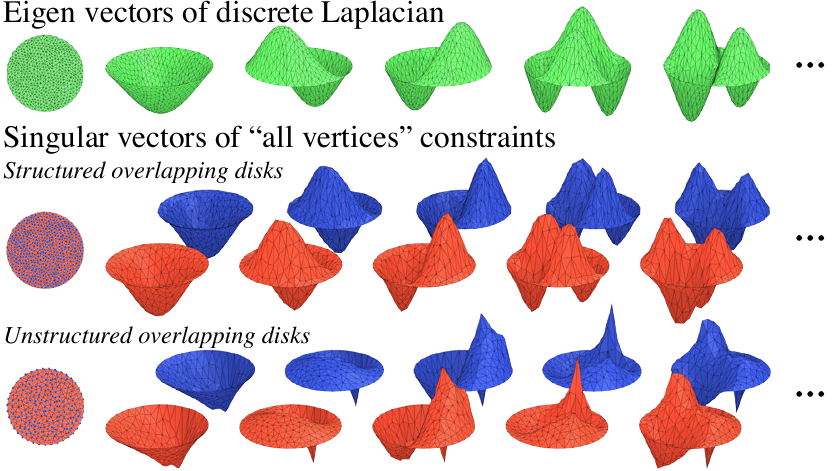}
  \caption{\label{fig:eigen-modes} 
  Eigen modes of the constraint matrix built from fixing all vertices of two
  overlapping, non-matching disk meshes resemble those of the discrete
  Laplacian.
  More irregular meshing produces less smooth modes.
  }
  \vspace*{-0.23cm}
\end{figure}

This is not a coincidence. The constraint matrix $\C$ in
\refequ{constraint-matrix} satisfies many desired properties (constant
precision, linear precision, the maximum principle and local support; see
\cite{Wardetzky:2007:DLO}) of a \emph{discrete Laplacian} on the ``joint mesh''
over $Ω₁∩Ω₂$ created by connecting each vertex of $Ω₁$ to the vertices of its
containing simplex containing in $Ω₂$ and \emph{vice-versa}.
Performing eigen analysis on $\C$ reveals that it responds as a discrete
operator strikingly similarly to the FEM discrete (cotangent) Laplacian (see
\reffig{eigen-modes}).
Due to this relationship, we call the artificial stiffening due to constraining
all overlapping vertices \emph{harmonic locking}.

We will defer our discussion of attempting to soften this equality constraint
to \refsec{results} and instead return to the smooth setting to derive a
locking-free solution from first principles.

\section{Boundary-Only Coupling}
\label{sec:boundary-only}
The root of the locking troubles is the point-wise equality constraint
over the overlapping region $Ω₁∩Ω₂$ in \refequ{smooth-pointwise-constraint}.
Surely coupling is crucial.
If we remove this constraint entirely, then $u₁$ and $u₂$ will solve independent
Poisson equations, subject to emergent natural boundary conditions (in this
case, $∇u⋅\mathbf{n} = 0$) on the overlap boundary $∂(Ω₁∩Ω₂)$.  In other words,
these zero normal derivative boundary conditions uniquely
determine $u₁$ and $u₂$.

The fact that minimizers of our energy in \refequ{smooth-dirichlet-energy}
are uniquely determined by boundary conditions can be spun to play in our favor
when searching for non-locking coupling constraints.
Concretely, we will now show that it is sufficient to restrict the pointwise
equality constraints from the entire intersection region $Ω₁∩Ω₂$ in
\refequ{smooth-pointwise-constraint} to \emph{only} its boundary $∂(Ω₁∩Ω₂)$:
\begin{align}
  \label{equ:smooth-boundary-constraint}
  u₁(\x) & = u₂(\x) & ∀ \x ∈ ∂(Ω₁∩Ω₂).
\end{align}

We must show that minimizing the deconstructed energy in
\refequ{smooth-deconstructed-energy} over $u₁$ and $u₂$ with this constraint
instead of \refequ{smooth-pointwise-constraint} remains equivalent to
the minimization over $u$ in \refequ{smooth-dirichlet-energy}.

Assume that \newhl{$u₁$ and $u₂$ are minimizers of
\refequ{smooth-deconstructed-energy} satisfying $u₁ = u₂ |_{Ω₁ ∩ Ω₂}$}, then by uniqueness of energy
minimizers and equivalence with the energy in \refequ{smooth-dirichlet-energy},
$u=u₁ |_{Ω₁}$ and $u=u₂ |_{Ω₂}$.
We must show that minimizing \refequ{smooth-deconstructed-energy}
implies that $u₁ = u₂ |_{Ω₁ ∩ Ω₂}$.

Given minimizers $u₁$ and $u₂$ of \refequ{smooth-deconstructed-energy}, let us
define $u₁ = u₂ := h |_{∂(Ω₁ ∩ Ω₂)}$. It does not matter that we do not
explicitly know the value of $h$. It is enough that it is well defined
implicitly by solving the problem in \refequ{smooth-deconstructed-energy}
subject to \refequ{smooth-boundary-constraint}.
Since the minimizers $u₁$ and $u₂$ satisfy the Dirichlet conditions on
their respective boundaries
(Equations~(\ref{equ:smooth-dirichlet-1}-\ref{equ:smooth-dirichlet-2})), we can
add the following constraints to \refequ{smooth-deconstructed-energy}  without
changing the minimum:
\begin{align}
  u₁(\x) = u₂(\x) &= h(\x) & ∀ \x ∈ ∂(Ω₁ ∩ Ω₂).
\end{align}
Minimizers to our quadratic energy are uniquely determined by the values on the
boundary of the domain, so we can isolate the problem for the
overlapping region $Ω₁ ∩ Ω₂$, for example:
\begin{align}
  \min_{u₁}           & \tfrac{1}{2}∫_{Ω₁ ∩ Ω₂} \tfrac{1}{2}\|∇u₁\|² - u₁ \ dA \\
  \text{ subject to } & u₁(\x) = h(\x) & ∀ \x ∈ ∂(Ω₁ ∩ Ω₂),
\end{align}
whose optimal argument is identical to the analogous problem replacing $u₁$ with
$u₂$, thus implying that the two functions agree on the overlapping region:
$u₁ = u₂ |_{Ω₁ ∩ Ω₂}$.

Schwarz noticed this over a century ago \shortcite{Schwarz1870}. Since then, it has
been been exploited for domain decomposition for parallelization and memory
decoupling for iterative solvers discussed in \refsec{related}.

Analogous to the enforcement of Dirichlet boundary conditions, in the discrete
linear FEM setting, we constrain only boundary vertices of $Ω₁$ lying inside the other
domain $Ω₂$ or \emph{vice-versa}:
\begin{align}
  \label{equ:boundary-only-constraints}
  u_{1i} &= \sum\limits_{j=1}^{n₂} u_{2j} φ_{2j}(\vv_{1i}) & ∀  i \text{ such that } \vv_{1i} ∈ ∂Ω₁∩Ω₂, \\
  u_{2j} &= \sum\limits_{i=1}^{n₁} u_{1i} φ_{1i}(\vv_{2j}) & ∀  j \text{ such that } \vv_{2j} ∈ ∂Ω₂∩Ω₁.
\end{align}
These constraints are a subset of the rows of $\C$ in
\refequ{naive-constraints},
and we call this much smaller matrix $\A ∈ \R^{(b₁+b₂) × (n₁+n₂)}$,
where the mesh of $Ω_i$ has $b_i$ overlap-boundary vertices.

\begin{figure}
\includegraphics[width=\linewidth]{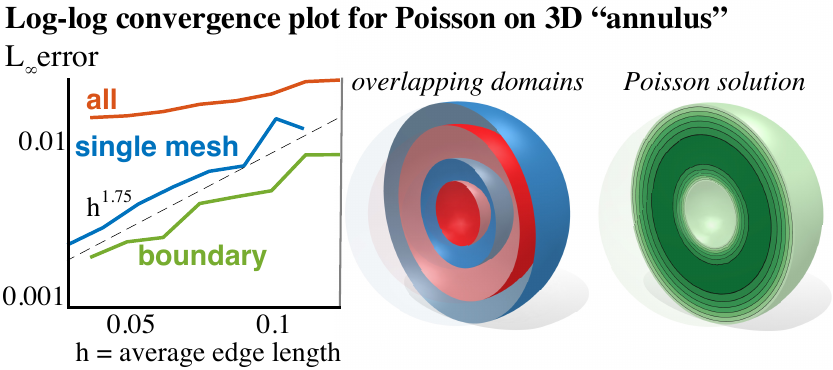}
  \caption{\label{fig:3d-annulus-convergence}
    \newhl{Constraining \emph{all} vertices between two concentric
    spherical shells (``3D annuli''; cut view in middle) does not
    appear to converge for the Poisson equation $∆u = 1$ (right), while
    \emph{boundary only} exhibits similar convergence to a non-overlapping
    \emph{single mesh}.}
  }
\end{figure}

Not only does fixing the boundary result in a smaller number of constraints and
thus typically a better conditioned system, but also the discrete approximations
are free of locking artifacts. We see this immediately in the 1D example in
\reffig{1d-poisson}. The boundary constraints do not show up visible in the
constraint space when reproducing hyperbolic or parabolic functions
\reffig{saddle-vs-parabola}. In Figures~\ref{fig:2d-annulus-convergence}
and~\ref{fig:3d-annulus-convergence}, convergence with respect to mesh
resolution for second-order problems roughly matches that of using a single
unified mesh. Recall that we are purposely avoiding creating such a unified
mesh, especially in $\R³$, where mesh surgery and likely manual intervention and
parameter tuning would be necessary. Instead, complex shapes can be created by
overlapping many solid subdomains and coupling solutions using our proposed
boundary-only constraints.

\subsection{Multiple Overlapping Subdomains}
In general, a complex shape may be composed of the union of $K>1$ subdomains: 
\begin{equation}
  Ω = \bigcup\limits_{i=1}^K Ω_i.
\end{equation}
All of our derivations so far for $K=2$ extend easily to $K>2$.
Our deconstructed energy has the form
\begin{align}
  \label{equ:multi-energy}
  \sum\limits_{i=1} ∫_Ω \frac{1}{∑_{j=1}^K χ_j } \left(\tfrac{1}{2}\|∇u_i\|² - u_i\right)\dA,
\end{align}
where $χ_j$ is the characteristic function of $Ω_j$ (i.e., $χ_j(\x) = 1$ for $\x ∈ Ω_j$ and $=0$
otherwise). \newhl{We defer the implementation details and matrix construction
to \refapp{details}.}

Many problems in geometry processing are slight variations on the minimization
of this energy. For example, implicit time-integrations of the wave equation
replace the unit potential with acceleration, while the heat equation replaces
this with a temperature field \cite{SwissArmyKnife:2014}. While these changes to
the basic Poisson solver here are nominal and left to the reader, increasing the
differential order of the energy requires specific attention (see
\refsec{higher-order}). \newhl{Before this, we discuss two important
considerations during discretization.}

\subsection{\newhl{Quadrature}}
\label{sec:quadrature}
\begin{figure}
  \includegraphics[width=\linewidth]{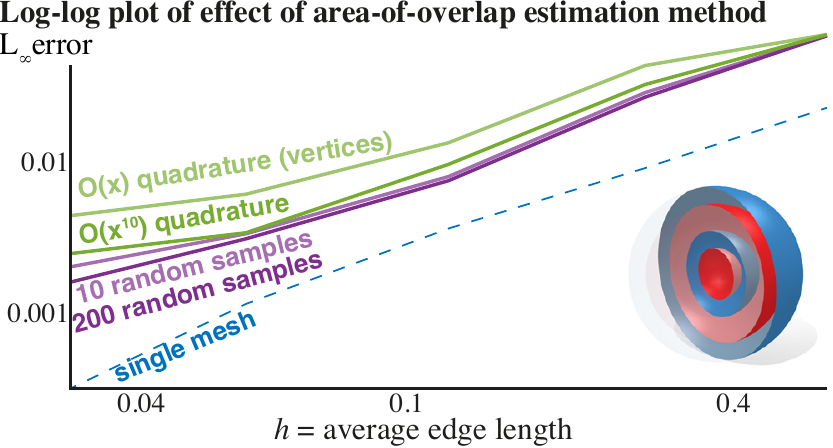}
  \caption{\label{fig:quad-vs-random} 
  \newhl{Quadrature for estimating partial volume at boundary elements can increase
  convergence, with diminishing returns.}
  }
\end{figure}

\newhl{When discretizing the integral in \refequ{multi-energy}, we must 
approximate the partial volume of tetrahedra
straddling the overlap boundaries. We compared various strategies.}
We specifically \emph{avoid} computing this analytically or splitting elements
as this is tantamount to the mesh boolean problem and would inherit its
numerical challenges and robustness issues.
Instead, we observe that numerical quadrature or Monte Carlo sampling
improves accuracy and indeed help convergence, albeit with diminishing returns
(see \reffig{quad-vs-random}).
\newhl{Approximating this integral is simpler than remeshing. We avoid computing
exact intersections or new combinatorics.}
  
%
Unless otherwise noted, we simply treat an element as fractionally
inside or outside another mesh by averaging the number of domains each corner
positions lies within (i.e., first-order quadrature).

\begin{figure}
\includegraphics[width=\linewidth]{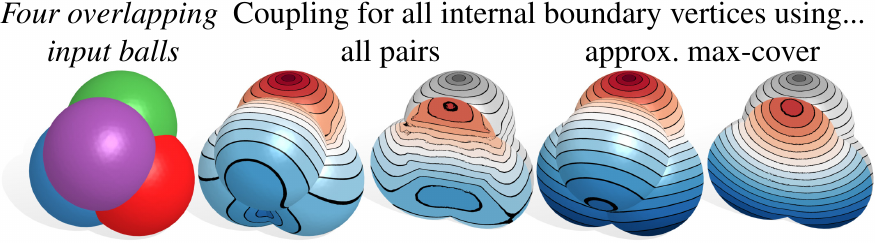}
  \caption{\label{fig:4-spheres-geodesics-in-heat}
  Boundary coupling across all subdomain pairs leads to messy locking near the
  overlap boundary \newhl{(6001 constraints)}.  Instead, our heuristic keeps exactly
  one constraint per overlap boundary vertex
  \newhl{(3893 constraints)}.}
\end{figure}

\subsubsection{\newhl{Constraint Thinning}}
\label{sec:constraint-thinning}

The simplest way to extend our boundary coupling constraints for $K=2$ in
\refequ{boundary-only-constraints} is to consider all possible pairs of
the $K$ subdomains:
\begin{align}
  \label{equ:multi-constraints}
  u_{ai} &= \sum\limits_{j=1}^{n_b} u_{bj} φ_{bj}(\vv_{ai}) \ \ \ ∀ i,a≠b \text{ such
  that } \vv_{ai} ∈ ∂Ω_a∩Ω_b.
\end{align}

\begin{figure*}
  \includegraphics[width=\linewidth]{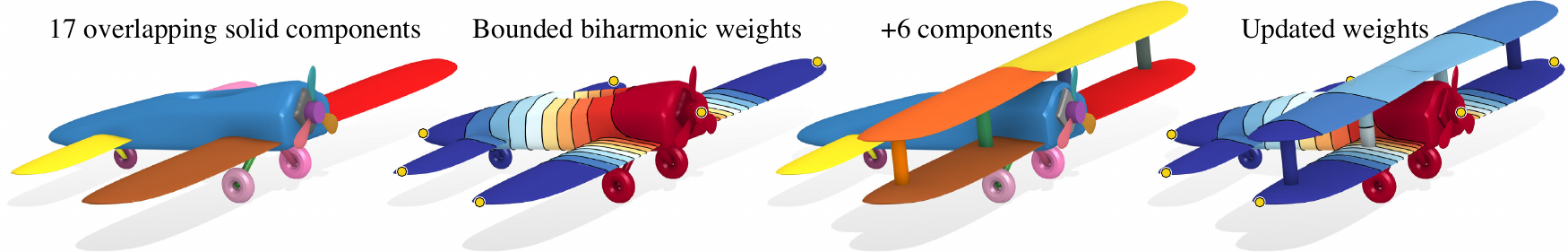}
  \caption{\label{fig:wwi-plane} Updating the shape with another stack of wings
  does not require updating a unified tet mesh. Instead new components are
  tet-meshed independently and added to the system algebraically.}
\end{figure*}

For shapes where many subdomains overlap on the same region, a boundary vertex
of one subdomain may show up in $>1$ other subdomains, resulting in equality
constraints for each. This unnecessarily reduces
the search space and tarnishes the solution near the overlap boundary (see
\reffig{4-spheres-geodesics-in-heat}).
\newhl{Much like in \cite{PTSZ11}, the transitivity of the equality above makes
it so that} we only \emph{need} one coupling constraint for each boundary
vertex. \newhl{We cannot be satisfied with finding
\emph{any} maximal spanning tree of constraints since that may still concentrate
constraints near a single vertex.}


\begin{figure}
\includegraphics[width=\linewidth]{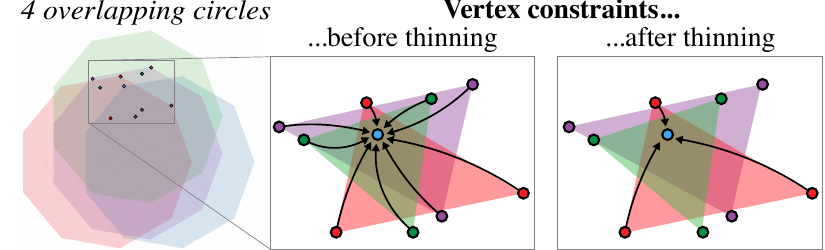}
  \caption{\label{fig:thinning}
\newhl{Pair-wise constraints on multiple overlapping subdomains lead to 
redundancy, which we remove via constraint thinning.
The highlighted vertex of the blue mesh in this didactic example 
 is interpolated by elements in the green, red and purple meshes, 
when only one of these would be necessary.}}
\end{figure}

We experimented with various heuristics for picking which constraint to keep
\newhl{for each fixed vertex}.  \newhl{Removing all but} the first constraint
creates a slight bias to the arbitrary ordering of the domains. Selecting a
random constraint works reasonably well, but still results in many vertices
involved in multiple constraints.
Averaging or softening constraints also helps, but increases complexity.

Ideally we would like to maximize the total number of vertices involved in the
constraints (to diffuse the constraints) while still ensuring exactly one
constraint per overlap-boundary vertex.
Viewing the constraint matrix $\A$ as graph, this selection is a form of vertex
cover problem.
%

\newhl{We approximate the maximum cover by
\emph{scoring} vertices based on how many 
constraints they are involved in; similarly,
we score each constraint by averaging the scores of the vertices involved. 
For each vertex involved in more than one constraint, we keep the least
saturated (lowest scored) constraint only and remove the rest, as shown in
\reffig{thinning}.}
%
\emph{Thinning} the constraints in this way significantly helps avoid issues
near boundaries when multiple shapes overlap
(see \reffig{4-spheres-geodesics-in-heat}).
\newhl{The decrease in the number of constraints also reduces the linear system
size, albeit with marginal affect on performance.}

\section{Higher-Order Partial Differential Equations}
\label{sec:higher-order}
Methods in geometry processing often go beyond second-order PDEs to model
problems requiring smoother continuity at constraints
\cite{Botsch:2004:AIF,LaplacianMeshEditing:2004,Jacobson:MixedFEM:2010,Stein2017}
or higher-order control \cite{Finch:2011:FVG,Joshi:2008:RAI} (see
\reffig{wwi-plane}). Returning briefly to the smooth setting, we focus on the
squared Laplacian energy to extend our consideration of deconstructed domains to
higher-order PDEs:
\begin{equation}
  \label{equ:smooth-squared-laplacian}
  \min_u ∫_Ω (∆u)² \dA,
\end{equation}
resulting in the fourth-order bi-Harmonic equation:
\begin{align}
  \label{equ:smooth-bilaplace}
  ∆²u(\x) &= 0 & ∀ \x ∈ Ω.
\end{align}

The second derivatives of this energy are not immediately discretizable
using linear FEM, so we introduce an auxiliary function $z$ and solve
the equivalent constrained minimization problem:
\begin{align}
  \label{equ:smooth-aux}
  \min_{u,z}         & ∫_Ω z² \dA, \\
  \text{subject to } & ∆u(\x) = z(\x) & ∀  \x ∈ Ω.
\end{align}
Applying the Lagrange multiplier method and Green's identity, this transforms
into a saddle problem involving only first derivatives:
\begin{equation}
  \mathop{\text{saddle}}_{u,z,μ} ∫_Ω \left( z² + ∇μ⋅∇u + μz \right) \dA + \text{ boundary terms}
\end{equation}
where $μ$ is the Lagrange multiplier function and we defer discussion of
boundary terms to previous works (e.g., \cite{Stein2017}).

\begin{figure*}
  \includegraphics[width=\linewidth]{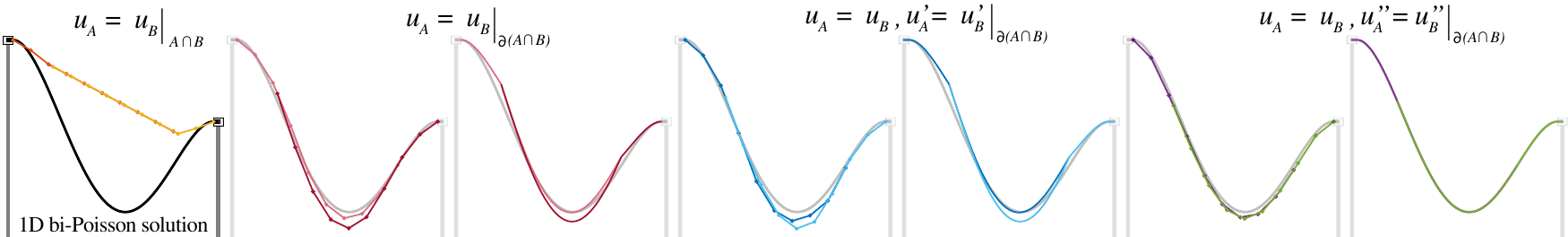}
  \caption{
    \label{fig:1d-bipoisson-locking}
    Enforcing equality at \emph{all vertices} of two overlap domains results in
    locking (red \& yellow).
    For higher-order PDEs, coupling \emph{by value alone} at overlap
    boundaries avoids extreme locking, but resolution refinement
    reveals non-smoothnesses (pinks).
    Attempting to enforce derivative continuity by coupling values and first
    derivatives results in \emph{local locking} and non-smoothness persists
    (blues).
    We couple primary values \emph{and} auxiliary values constrained to the
    Laplacian ($u''$ in 1D) at sub-domain boundaries: the solution is smooth
    (purple \& green).
  }
\end{figure*}

We now have a problem involving only first derivatives which we can discretize
using \newhl{multiple sets of linear finite elements (i.e., mixed FEM)}. After
factoring out $μ$, the resulting system has the symmetric matrix form of a KKT
system:
\begin{equation}
  \label{equ:mixed-KKT}
  \left(\begin{array}{cc}
    \mathbf{0} & \L^\transpose \\
    \L         & -\M
  \end{array}\right)
  \left(\begin{array}{c}
    \u \\
    \z
  \end{array}\right)
  =
  \left(\begin{array}{c}
    \mathbf{0} \\
    \mathbf{0} 
  \end{array}\right)
\end{equation}

\subsection{\newhl{Unsuccessful Low-Order Boundary-Only Coupling}}
\label{sec:low-order}
While the second-order Poisson equation requires one set of boundary conditions (e.g.,
fixed values or fixed normal derivatives), the fourth-order bi-Laplace equation
in \refequ{smooth-bilaplace} requires two sets of boundary conditions to
identify a unique solution. For example, we can fix both the value \emph{and}
the normal derivative along the boundary $∂Ω$ (i.e., fix low-order quantities).
If we explicitly fix \emph{only} the value along the boundary when minimizing
the squared Laplacian energy, then \emph{natural
boundary conditions} will emerge to ensure uniqueness (cf.\ \cite{Stein2017}).
This also occurs in the mixed FEM discretization.
Fixing only the value along the overlapping region for two
subdomains $Ω₁ ∩ Ω₂$ couples the function values together, but produces a
noticeable ``kink'' (see \reffig{1d-bipoisson-locking}).
We are witnessing the natural boundary conditions on one subdomain's function
(in this case $∆u=0$) disagreeing with the derivatives of \emph{other}
subdomain's function: i.e., in general, $∆u₁ = 0 ≠ ∆u₂$ on $∂Ω₁ ∩ Ω₂$.

To take advantage of the same uniqueness properties used in
\refsec{boundary-only}, we must ensure that each function is sufficiently
constrained with boundary conditions. One idea would be to trivially extend our
boundary-only coupling by fixing the value \emph{and} normal derivative along
the overlapping boundary:
\begin{align}
  u₁(\x) &= u₂(\x)                 & ∀ \x ∈ ∂(Ω₁∩Ω₂),\\
  ∇u₁(\x)⋅\n(\x) &= ∇u₂(\x)⋅\n(\x) & ∀ \x ∈ ∂(Ω₁∩Ω₂),
\end{align}
where $\n(\x)$ is the normal vector pointing outward from the overlapping region
$Ω₁∩Ω₂$.
In the smooth setting, we can quickly confirm that this is equivalent to the
original energy minimization problem in \refequ{smooth-squared-laplacian}
following the same reasoning in \refsec{boundary-only}.

These \emph{low-order} coupling constraints are simple to discretize using
linear FEM, but unfortunately do not lead to a convergent system. Fixing
directional derivatives across the two functions leads to harmonic locking
locally (the one-ring of vertices at the overlapping region boundary). This
region shrinks with mesh refinement, but the problem persists: effectively the
solution locks so that natural boundary conditions emerge, albeit one-ring into
the overlapping domain (see \reffig{1d-bipoisson-locking}).

\subsection{Higher-Order Boundary-Only Coupling}
\label{sec:higher-order-boundary}
Fortunately, the bi-Laplace equation in \refequ{smooth-bilaplace} is also uniquely
determined by other combinations of boundary conditions. Such combinations of
low- and high-order conditions sometimes appear directly during problem modeling
(e.g., \cite{Joshi:2008:RAI}). The introduction of the auxiliary variable $z =
∆u$ in \refequ{smooth-aux}, makes the choice of fixing the value and the Laplacian of $u$ along the
boundary particularly easy to describe:
\begin{align}
  u₁(\x) &= u₂(\x)  & ∀ \x ∈ ∂(Ω₁∩Ω₂),  \\
  z₁(\x) &= z₂(\x)  & ∀ \x ∈ ∂(Ω₁∩Ω₂).
\end{align}
During discretization using mixed FEM, we add these constraints to the
Lagrangian's KKT system in \refequ{mixed-KKT} directly, resulting in a larger
KKT system:
\begin{equation}
  \label{equ:mixed-KKT-coupled}
  \left(\begin{array}{cccc}
    \mathbf{0} & \L^\transpose & \A^\transpose & \mathbf{0} \\
    \L         & -\M           & \mathbf{0}    & \A^\transpose \\
    \A         & \mathbf{0}    & \mathbf{0}    & \mathbf{0}    \\
    \mathbf{0} & \A            & \mathbf{0}    & \mathbf{0}    \\
  \end{array}\right)
  \left(\begin{array}{c}
    \u \\
    \z \\
    λ_u \\
    λ_z
  \end{array}\right)
  =
  \left(\begin{array}{c}
    \mathbf{0} \\
    \mathbf{0} \\
    \mathbf{0} \\
    \mathbf{0} \\
  \end{array}\right),
\end{equation}
where $λ_u,λ_z ∈ \R^{n₁+n₂}$ are vectors of Lagrange multipliers enforcing
boundary coupling constraints on $\u$ and $\z$ respectively, \newhl{and $\A$ is the linear constraint matrix}.
While the constraints on the values in $\u$ are straightforward, the
constraints on the auxiliary values $\z$ may be interpreted as acting
orthogonally to the original mixed FEM constraint that $\M \z = \L u$.

\begin{figure}
  \includegraphics[width=\linewidth]{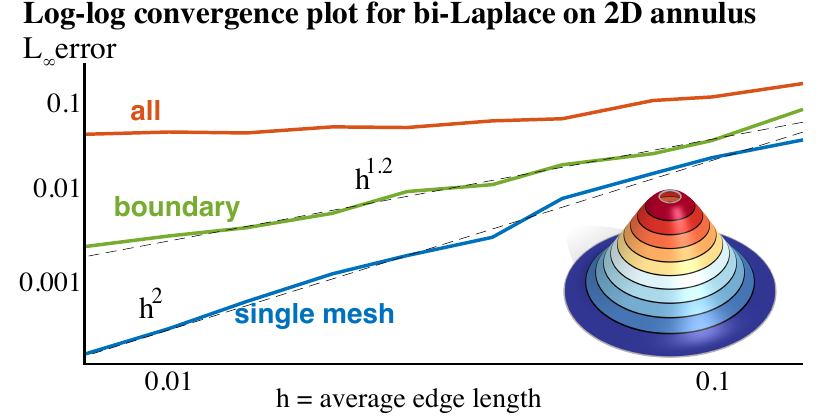}
  \caption{
  \label{fig:2d-bilaplace}
  \newhl{The behavior of our different sets of constraints for the bi-Laplace
  equation $\Delta^2 u =0$ mirrors that of the Poisson equation solver shown
  previously.}
  }
\end{figure}

This discretization avoids the ``kink'' of the low-order boundary
coupling in \refsec{low-order} (see \reffig{1d-bipoisson-locking}). We see
convergence with respect to mesh refinement (see \newhl{\reffig{2d-bilaplace}}).

A remaining issue with our discretization is that mixed FEM results in a saddle
problem, rather than a standard convex, linearly constrained quadratic energy
minimization. \newhl{However, this is overcome by rearranging terms
algebraically (see \refapp{massage}).}


\section{Experiments \& Results}
\label{sec:results}
We have implemented our method using \textsc{Matlab} using finite element
operators from \textsc{gptoolbox} \cite{gptoolbox} and point location routines
from \textsc{libigl} \cite{libigl}.
We use \textsc{TetGen} \cite{tetgen} to mesh the subdomains in all examples
except the sphere and 3D annulus
test cases, for which we use \textsc{Quartet} \cite{Doran:2013}.
We use \textsc{Triangle} \cite{shewchuk96b} for 2D meshing.
On our MacBook Pro with a 3.5GHz Intel Core i7 with 16 GB of memory, the
performance bottleneck is always the linear solve (\textsc{Matlab}'s
\texttt{ldl}), eigen decomposition (\textsc{Matlab}'s \texttt{eigs}) or
quadratic programming optimization (\textsc{Mosek}'s \texttt{quadprog}).
\newhl{For completeness, we list runtime performance in
\reftab{timing-examples}}.

\begin{table}
\centering
\begingroup
\newcolumntype{$}{>{\global\let\currentrowstyle\relax}}
\newcolumntype{^}{>{\currentrowstyle}}
\newcommand{\rowstyle}[1]{\gdef\currentrowstyle{#1}%
  #1\ignorespaces
}
\newcolumntype{P}[1]{>{\centering\arraybackslash}p{#1}}
\ra{1.2}
\setlength{\tabcolsep}{5.5pt}
\rowcolors{2}{white}{lightbluishgrey}
\begin{tabular}{l r r r r r r}
\rowcolor{white}
\toprule
  Shape      &        \#Tets  &     $K$ &   Build $\A$ & Problem & Solve \\
\midrule
  Android    &       113118   &      33 &       0.56 s & BBW      &  9.06 s\\
  Bug        &       159533   &      16 &       0.47 s & MSBK     & 21.73 s \\
  Jet        &       226548   &      14 &       0.48 s  & Eigen   &  1.91 s \\
  Bi-Plane   &       321237   &      23 &       1.05 s & BBW      & 37.61 s \\
  Microscope &       348099   &      24 &       1.09 s & Heat     &  4.84 s \\
  Pistol     &       412798   &      18 &       1.08 s & Wave     &  4.70 s  \\
  Alien      &       682399   &      32 &       2.57 s & Geodesic & 13.45 s \\
\bottomrule
\end{tabular}
\endgroup
  \caption{\newhl{
    Performance timings: \#Tets is the total number of tetrahedra across the $K$
    overlapping components. We list the
    runtime for constructing the constraints matrix (\emph{Build} $\A$)  and
    then conducting the resulting constrained (example-dependent) optimization
    (\emph{Solve}).}}
\label{tab:timing-examples}
\end{table}
\begin{figure}
  \vspace*{-0.2cm}
\includegraphics[width=\linewidth]{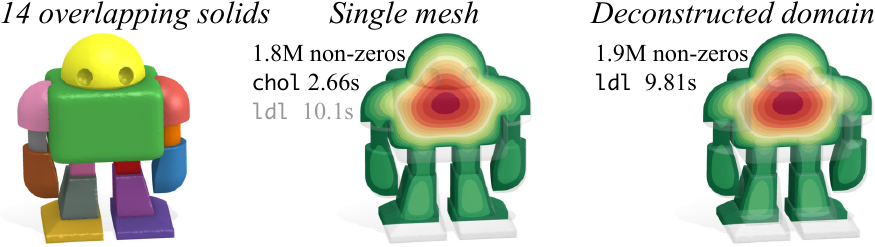}
\caption{
  \label{fig:robot-vs-single}
  \newhl{
    Traditional mesh-union then tetrahedralization succeeds on this example,
    enabling an \emph{ad hoc} performance comparison. System matrices for
    solving the Poisson equation 
    require similar memory, however our \textsc{MATLAB} implementation uses
    $LDLT$ on our resulting Lagrangian: about $3.7×$ slower than Cholesky,
    here.}}
  \vspace*{-0.5cm}
\end{figure}

\newhl{
  While our main focus is to improve robustness, we observe systematically
  predictable trends in the runtime performance.
  For example, consider solving a Poisson equation on a \emph{single mesh}
  of a solid domain with $O(n³)$ vertices. For a typical FEM-quality mesh, the
  performance will be determined by performing a linear system solve on a sparse
  matrix with $O(n³)$ non-zeros. In the absence of other constraints, this
  matrix will be positive definite, affording Cholesky decomposition.
  For our deconstruction of the same domain into $K$ overlapping components and
  $O(n³)$ total vertices across all meshes, we build the \emph{boundary-only}
  constraints matrix $\A$ which (under mild assumptions) will contain $O(Kn²)$
  non-zeros.
  In contrast, fixing \emph{all vertices} in the overlapping region would
  require $O(Kn³)$ non-zeros.
  Using, e.g., the Lagrange multiplier method to enforce our constraints results
  in an indefinite sparse system matrix with $O(n³ + K n²)$ non-zeros (solved,
  e.g., with $LDLT$-decomposition).
  In practice, $K$ is often quite small and the difference in performance
  between solving on a \emph{single mesh} and a deconstructed domain boils down
  to the performance of sparse Cholesky versus sparse $LDLT$-decomposition ---
  with the important caveat that solving on a \emph{single mesh} is often
  impossible without user-intervention.
  %
  In \reffig{robot-vs-single}, we found an example where mesh-union followed by
  tetrahedralization \emph{does} create a useable mesh:
  Cholesky for the single mesh is roughly $3.7×$ faster than $LDLT$ on our
  constrained system.
  As future work, it would be interesting to further exploit our deconstructed
  domains for performance acceleration and parallelization (see, e.g.,
  \cite{Loehner}).
}

\begin{figure}
\includegraphics[width=\linewidth]{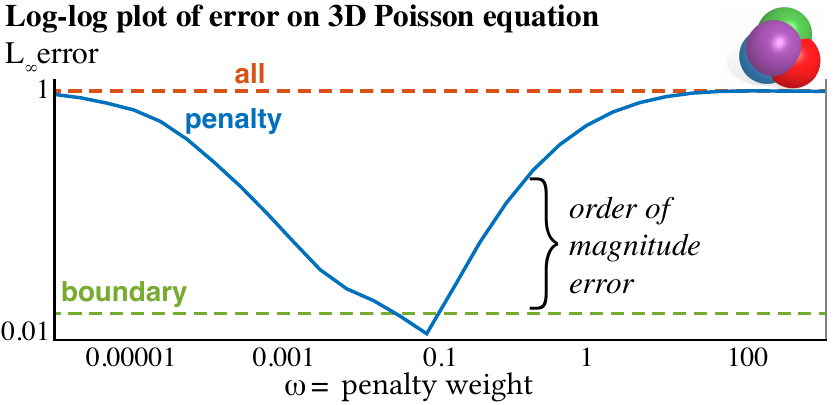}
  \caption{\label{fig:least-squares-comparison}
  Weak constraints are sensitive to the penalty weight. There exists a good value
  (here, $ω=0.1$), but finding this is non-trivial and problem dependent. An
  incorrect choice can be disastrous.
  Our \emph{boundary} coupling is parameterless and achieves low error.}
\includegraphics[width=\linewidth]{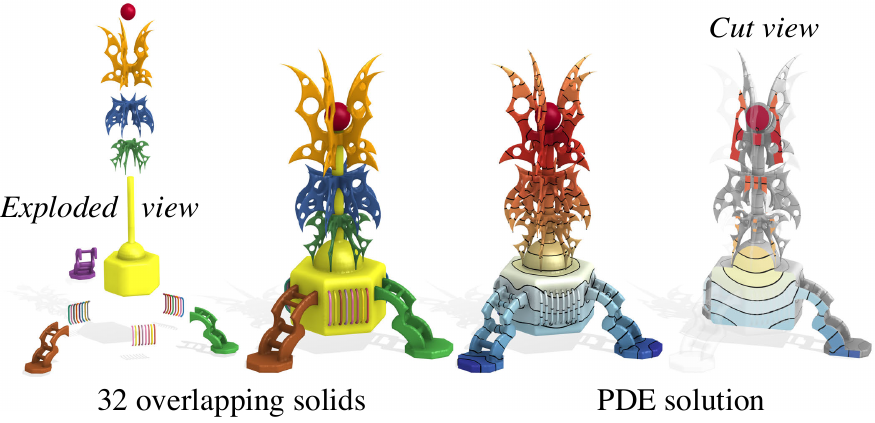}
\caption{
  \label{fig:teaser}
  \newhl{Following the steps described in \cite{Crane:GIH:2012}, we can use our method for computing geodesic distances on a complex shape without a tetrahedral discretization of the domain.}
}
\end{figure}

\begin{figure*}
  \includegraphics[width=\linewidth]{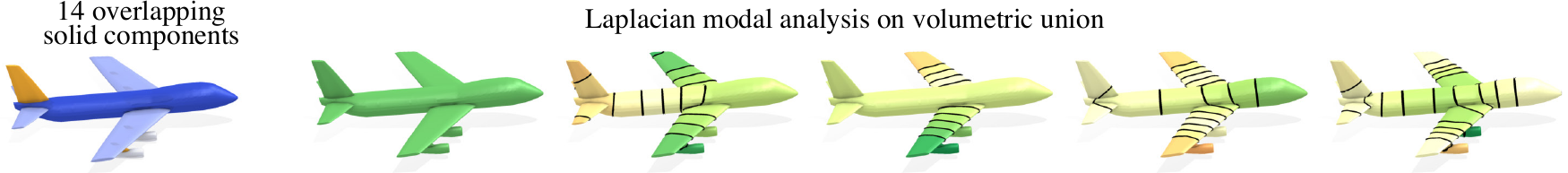}%
  \vspace*{-0.2cm}%
  \caption{\label{fig:plane}
  Our general method can be applied to a variety of problems involving
  discrete differential geometry including eigen analysis.}
  \vspace*{-0.7cm}
\end{figure*}
\begin{figure}
  \includegraphics[width=\linewidth]{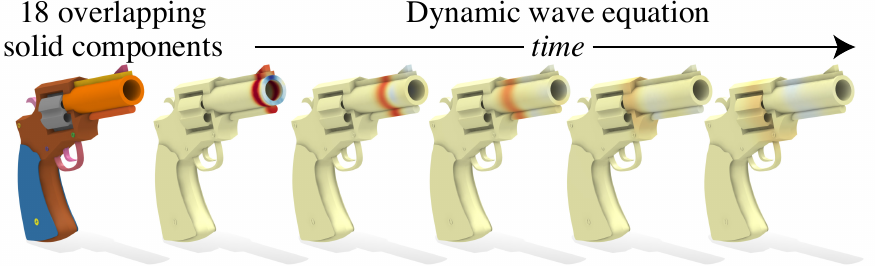}
  \caption{\label{fig:pistol}
  Our method is accurate enough to repetitive solves for dynamics, such as this
  shock wave equation simulation.
  }
\end{figure}
\newhl{An alternative to our  boundary-only \emph{hard} constraints would be to
enforce \emph{weak} constraints at all vertices in the overlapping regions.}
In \reffig{least-squares-comparison}, we show that, yes, weak constraints can
work, but one must choose the penalty weight carefully. \newhl{In this
experiment, if the penalty is
too weak the solutions on different subdomains
become decoupled;
too strong the solution locks up just as much
as the strong constraints.}
This is not a situation where a different constraint handler will help.
For example, the Augmented Lagrangian or Alternating Direction Method of
Multipliers (ADMM) methods are numerical techniques for effectively driving the
penalty weight to infinity, but in this limit the solution is simply the locked
up solution.
Meanwhile, the ``correct'' penalty weight will depend on the mesh resolution,
constraint constellation and solution. 
This may vary spatially: a good weight here may cause locking over there.

\begin{figure}
  \includegraphics[width=\linewidth]{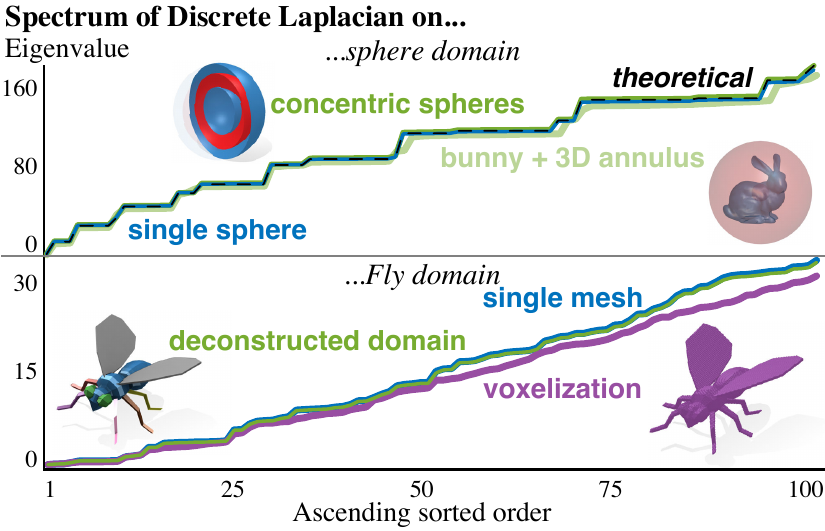}%
  \vspace*{-0.06cm}%
  \caption{\label{fig:eigenanalysis}
\newhl{The spectral behaviour of our Laplacian operator constructed
 using solely the information from the primitive's meshes (green lines) 
 approaches the analytical spectra (left) in the same way as that obtained
  from traditional FEM on a unified mesh of the domain (left and right, 
  blue line).
  }
}
  \includegraphics[width=\linewidth]{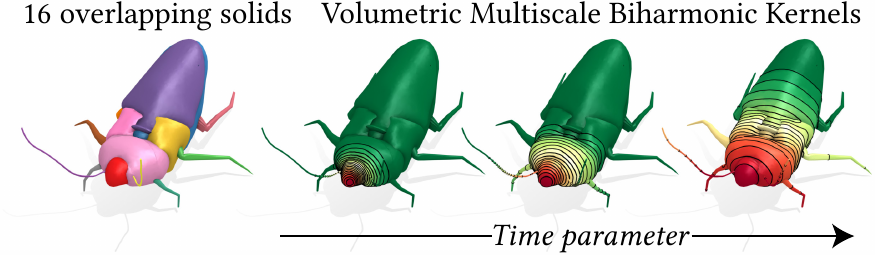}%
  \vspace*{-0.06cm}%
  \caption{\label{fig:bug-msbk}
  Our high-order boundary coupling constraints complement advanced biharmonic
  energy-minimization methods and additional constraints such as $L₁$ sparsity.}
  \includegraphics[width=\linewidth]{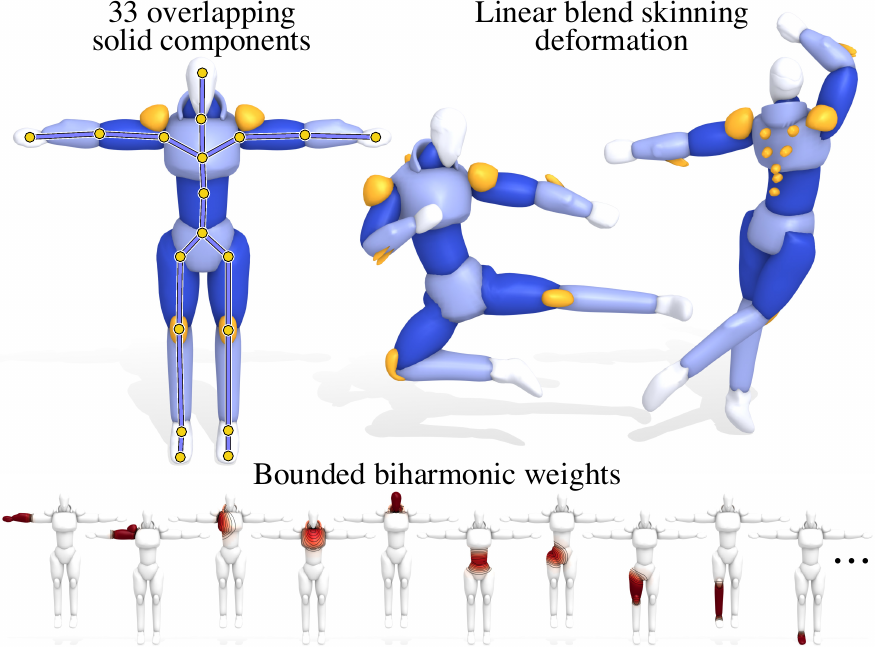}%
  \vspace*{-0.06cm}%
  \caption{\label{fig:typhoon-bbw}
  Minimizing the squared Laplacian subject to bound constraints produces
  automatic deformation bases: our method enables this over deconstructed
  domains.}
\end{figure}

We designed 2D and 3D convergence test scenarios (see
Figures~\ref{fig:2d-annulus-convergence} and \ref{fig:3d-annulus-convergence}).  We
compare $L_∞$ error to an analytic solution.
In \reffig{quad-vs-random}, we use the same setup to test partial area
estimation. For 10-point quadrature, we use the symmetric rules of Zhang et
al.~\shortcite{zhang2009set}.

\newhl{For irregular tetrahedral meshes, elements overlapping the boundary of
another domain typically contain multiple boundary vertices of that domain and
thus participate in multiple constraints (even in the simple overlapping 3D annuli
in \reffig{3d-annulus-convergence} \emph{involved} tetrahedra contain on average
3.18 boundary vertices).
For more complex shapes, the interior boundaries inherit the irregularity of the
overlapping parts. Our method does not smooth or alter these potentially
irregular boundaries (see \reffig{teaser}).}


We demonstrate the versatility of our constraints by expanding beyond the
Laplace ($∆u = 0$) and Poisson equations ($∆u = f$) to other equations found in
solid geometry processing.
In \reffig{microscope}, we demonstrate our boundary only constraints for solving
an implicit time step of the heat equation ($u - δt ∆u = u₀$). In
\reffig{teaser}, we solve the same heat equation \newhl{for $u$ then the Poisson
equation ${\Delta \varphi =- \Delta u /|\nabla u|}$  to approximate interior
distances $\varphi$} using the method of Crane et al.\ \cite{Crane:GIH:2012}.
In \reffig{pistol}, we visualize
shock wave through a pistol composed of many overlapping components 
($u - δt² ∆u = u₀ + δt \dot{u}₀$).

In \reffig{plane}, we use our boundary only constraints to conduct a
\newhl{Laplacian} modal analysis on a deconstructed domain. We enforce
constraints during eigen decomposition via the null space method
\cite{golub1973some}, but replace the QR decomposition with the sparser LUQ
decomposition.
\newhl{In \reffig{eigenanalysis}, we quantitatively validate our method using
the Laplacian spectrum. The smallest one hundred eigenvalues using our method
match the theoretical groundtruth for a sphere domain (and those computed using
standard linear FEM on a single mesh).
To extend this comparison to a more complex example where theoretical values are
not known, we found a shape where mesh-union followed by tetrahedralization
succeeds.
Compared to second-order finite differences over a high-resolution voxelization,
our spectrum better matches the spectrum found using a single unified mesh.
Higher-order elements --- known to improve spectral convergence
\cite{reuter2009discrete} --- could be used in either method, but do not affect
our main contribution of setting up constraints.}

By rearranging our higher-order coupling for bi-Laplacian problems in
\refsec{higher-order-boundary} into a convex energy minimization (see
\refapp{massage}) we can immediately implement advanced methods
involving $L₁$ sparsity inducing norms for shape descriptors, such as the
multiscale pre-biharmonic kernels \cite{Rustamov:2011} in \reffig{bug-msbk} and
inequality constraints such as the bounded biharmonic weights
\cite{Jacobson:BBW:2011}, used for real-time skinning deformations in
\reffig{typhoon-bbw}.
In \reffig{wwi-plane}, we demonstrate the robustness of our method to
large-scale geometry changes. Wings are added to the plane simply by overlapping
new solid components: we only need to tet-mesh the new components and add their
linear constraints to the system. In the classic geometry processing pipeline,
we would need to invoke mesh union and fragile global tet-meshing algorithms.
Our method avoids this.

\section{Limitations \& Future Work}
\label{sec:conclusion} 

We make a heavy assumption that the input domain is or can be deconstructed into
simple tetrahedralizable subdomains. While many models are originally created
using constructive solid geometry (CSG) operations, often only the (typically
poor triangle-quality) mesh-boolean result is available when it comes time to
solve a volumetric PDE. 
Therefore, we advocate to \emph{retain} these simpler domains and the
construction tree rather than preemptively resolving the mesh-boolean.
Nonetheless, our tetrahedralizers, \textsc{TetGen} \cite{tetgen} and
\textsc{Quartet}
\cite{Doran:2013}, still occasionally fail even on simpler subdomains.
To mitigate this we can preprocess problematic subdomains on a case-by-case
basis using \textsc{MeshFix} \cite{Attene:2010} and generalized winding numbers
\cite{barill2018}.
In this paper, we consider volumetric unions of polyhedral subdomains.
Other domains such as those modeled using metaball implicits \cite{Wyvill:1986}
or reconstructed from unstructured point clouds (e.g.,
\cite{PoissonSurfaceReconstruction06}) are not immediately suitable for our
method.
It is exciting to consider automatic methods for converting such domains into
unions of simpler primitives, perhaps with inspiration from advances in
approximate convex decomposition \cite{asafi2013weak}.

\begin{figure}
\includegraphics[width=\linewidth]{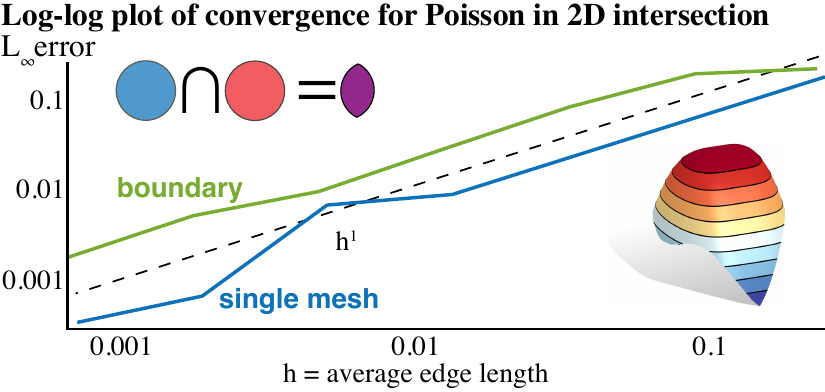}
\caption{
  \label{fig:csg}
   \newhl{Our method can be generalized to deal with other two-dimensional set operations, such as intersections and substractions. }
}
\end{figure}
While all examples presented in this paper deal exclusively with unions of
different shapes, one can conceive of certain variations that would make our
method valid for all CSG operations, such as intersections or
differences. We have promising initial results for intersecting two-domains in 2D (see \reffig{csg}) and are working
an extension to full 3D CSG trees.

\bibliographystyle{eg-alpha}
\bibliography{references}

\appendix

\section{\textbf{Deconstructed Domains Solver}}
\label{app:details}

\newhl{ This appendix provides a step-by-step construction of the discrete
solver for deconstructed domains.}

Without loss of generality let us assume three-dimensional domains ($d=3$).
The input to our method  is a set of $K$ overlapping, embedded, manifold
tetrahedral meshes with vertices $\{\V_1,…,\V_K\}$ so that $\V_i∈\R^{n_i × 3}$
contains the positions of the $i$th subdomain's $n_i$ vertices in its rows and
list of tetrahedral indices $\{\T_1,…,\T_K\}$  where the row-indices into $\V_i$
of the $i$th subdomain's $t_i$ tetrahedra appear as rows
$\T_i∈[1,\dots,n_i]^{t_i × 4}$.

We first build the constraint matrix 
$\A ∈ \R^{(b_1 + … + b_K) × (n_1 + … + n_K)}$, 
where $b_i$ are the number of boundary vertices of the $i$th mesh lying
inside a tetrahedron of \emph{any} other mesh.

Next we build the sparse discrete gradient matrix $\G_i ∈ \R^{3t_i × n_i}$ for
each domain and compute \emph{adjusted} volumes for each tetrahedron $\a_i ∈
\R^{t_i}$ (accounting for the $1/∑_{j=1}^K χ_j$ term in \refequ{multi-energy},
see \refsec{quadrature}).

From these we can construct a quadratic coefficients matrix (i.e., discrete
Laplacian) $\L_i ∈ \R^{n_i × n_i}$ for each domain:
\begin{equation}
  \L_i = \G_i^\transpose \diag{\a_i} \G_i,
\end{equation}
where $\diag{\x}$ for a vector $\x ∈\R^n$ creates $n×n$ matrix with $\x$ along
the diagonal.
We concatenate the contributions from each subdomain into a monolithic Laplacian
$\L∈\R^{(n_1 + … + n_K) × (n_1 + … + n_K)}$,
\begin{equation}
\L = \blkdiag{\L_1,…,\L_K},
\end{equation}
where $\blkdiag{\A,\B,…}$ creates a block diagonal matrix from matrices $\A$,
$\B$, …

Using the adjusted tetrahedral volumes in $\a_i$, we build a ``barycentric''
lumped diagonal mass matrix for each mesh $\M_i ∈ \R^{n_i × n_i}$ and stack
these as well to create the mass matrix of the entire system $\M∈\R^{(n_1 + … +
n_K) × (n_1 + … + n_K)}$,
\begin{align}
  (\M_{i})_{jj} &= ∑_{k ∈ N(j)} \tfrac{1}{4} (\a_i)_k \\
  \M       &= \blkdiag{\M_1,…,\M_K},
\end{align}
where $N(j)$ are the tetrahedra incident on vertex $j$.
Further accuracy could possibly be achieved by using a hybrid ``Voronoi'' mass
matrix \cite{Meyer:2003,Jacobson:MixedFEM:2010}.

Finally, we define $\u ∈ \R^{(n_1 + … + n_K)}$ as the vertically stacked vectors
of unknown per-vertex values across the $K$ subdomain meshes.

We may now pose the discretization of the energy minimization problem in
Equations (\ref{equ:multi-energy}-\ref{equ:multi-constraints}) using a standard
matrix form:
\begin{align}
           \min_{u} & ½ \u^\transpose \L \u - \u^\transpose \M \mathbf{1} \\
  \text{subject to } &    u_{ij} = g(\vv_{ij}) & ∀ \vv_{ij} ∈ ∂Ω ∩ ∂Ω_i  \\
  \text{       and } & \A \u = \mathbf{0}
\end{align}
where $\mathbf{1}$ and $\mathbf{0}$ are vectors ones and zeros respectively.
Vertices receiving boundary conditions or constraints are identified
combinatorially and located inside other meshes efficiently using a spatial
acceleration data structure (e.g., we use \textsc{libigl}'s AABB tree
\cite{libigl}), then \emph{thinned} by removing rows according to our
approximate max-cover criteria (see \refsec{constraint-thinning}).
We use the \textsc{Matlab} or \textsc{Mosek} quadratic programming solvers to
find an optimal $\u$.

\section{\textbf{Rearrangement into Quadratic Minimization}}
\label{app:massage}

A remaining issue with our discretization is that mixed FEM results in a saddle
problem, rather than a standard convex, linearly constrained quadratic energy
minimization. This means in practice we cannot send the system in
\refequ{mixed-KKT-coupled} to a standard quadratic programming solvers because
the top-left sub-block
\begin{equation}
  \label{equ:top-left}
  \left(\begin{array}{cc}
    \mathbf{0} & \L^\transpose \\
    \L         & -\M
  \end{array}\right)
\end{equation}
is not positive semi-definite.
However, we can resolve this by factoring out $\z = \M^{-1}(\L u + \A^\transpose
λ_z)$ resulting in the smaller KKT system:
\begin{equation}
  \left(\begin{array}{ccc}
    \L^\transpose \M^{-1} \L   & \L\M^{-1}\A^\transpose   & \A^\transpose  \\
    \A\M^{-1}\L^\transpose     & \A \M^{-1} \A^\transpose & \mathbf{0} \\
    \A                         & \mathbf{0}               & \mathbf{0}
  \end{array}\right)
  \left(\begin{array}{c}
    \u \\
    λ_z \\
    λ_u 
  \end{array}\right)
  =
  \left(\begin{array}{c}
    \mathbf{0} \\
    \mathbf{0} \\
    \mathbf{0} 
  \end{array}\right),
\end{equation}
where the top-left $2×2$ sub-block is now positive semi-definite. This system
arrives as the Euler-Lagrange equation for the constrained convex quadratic
minimization problem:
\begin{align}
  \min_{\u,λ_z}      & \| \M^{-\sfrac{1}{2}} ( \L \u + \A^\transpose λ_z)\|² \\
  \text{subject to } & \A \u = \mathbf{0}.
\end{align}
To avoid inverting the mass matrix and improve the conditioning of the objective
term, we introduce another auxiliary variable $\y ∈ \R^{n₁+n₂}$, arriving at the
final constrained problem in standard form:
\begin{align}
  \label{equ:bilaplace-convex}
  \min_{\u,λ_z,\y}  & \| \y \|² \\
  \text{subject to } & \A \u = \mathbf{0}, \\
  \text{and }        & \L \u + \A^\transpose λ_z = \M^{\sfrac{1}{2}}\y.
\end{align}

\end{document}